\documentclass{ws-p8-50x6-00}

\begin{document}

\title{Aspects of the Quantum Chromo Dynamics Phase Diagram}

\author{Francesco Sannino}

\address{NORDITA \\Blegdamsvej 17,
 Copenhagen \O, DK-2100 Denmark\\ E-mail:
francesco.sannino@nbi.dk}


\maketitle

\abstracts{I briefly review some aspects of the Quantum Chromo
Dynamics phase diagram at non zero temperature and quark chemical
potential. I then suggest new possible phases which can appear in
strongly interacting theories at non zero chemical potential.
Finally I describe the phase diagram as function of the number of
flavors and colors at zero quark chemical potential and zero
temperature.}

\section{Introduction}
There are already many excellent reviews on the QCD phase diagram,
its possible applications  and effective theories approaches
\cite{REV,{Satz:2002mg},{Shuryak:2001ik},{Halasz:1998qr},{Lombardo:2000rs},{Casalbuoni},{Nardulli}}.
Here I will first introduce some basic features of the phase
diagram and then suggest new phases which can affect it such as
vectorial-type condensation. I will also provide a brief summary
of a possible plan for the phase diagram when varying the number
of light flavors relative to the number of colors but keeping zero
temperature and quark chemical potential and motivate the
relevance for the physics beyond the standard model of particle
interactions.

\section{The Hot and Dense QCD Phase Diagram}

\subsection{Hadronic/Confining Phase}
At zero temperature and density we do not observe free quarks and
gluons. These states are permanently confined in hadronic
particles like pions, vectors, nucleons etc. In this regime  the
interactions among hadrons can be efficiently described via
effective Lagrangians built respecting the relevant symmetries of
the underlying theory.

\begin{figure}[ht]
\begin{center}
\begin{minipage}[c]{9cm}
\vspace*{-0.3cm} 
  \epsfxsize 9cm \epsfysize 9cm
  \epsfbox{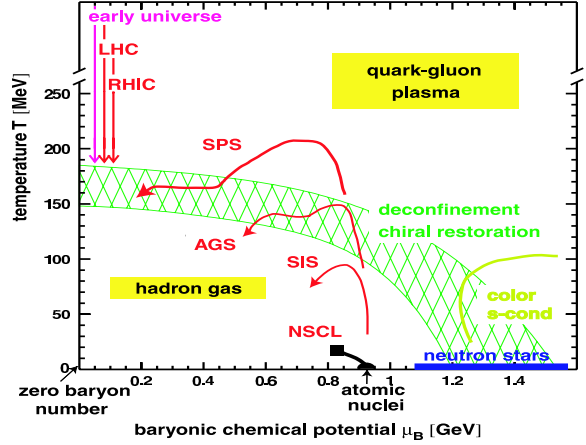}
\end{minipage}
\vspace*{-.5cm} \caption{Possible physical applications and region
of the QCD phase diagram explored by different experiments.}
\label{experiments}
\end{center}
\vspace*{-.5cm}
\end{figure}

\subsection{Hot and dilute QCD}

At temperatures just above the deconfinement phase transition  but
small quark chemical potential the theory enters a quark gluon
plasma phase (QGP) in which the underlying states are still
strongly interacting but deconfined. Only at asymptotically high
temperatures perturbative QCD calculations are really  reliable.
This phase is relevant for the physics of the early universe and
is accessible to earth based experiments like RHIC and LHC. Figure
1 \cite{Heinz:2001ax} shows the relevant experiments and possible
physical applications.

The study of the temperature driven confining-deconfining phase
transition has attracted much interest and work. Thanks to lattice
simulations we now have a great deal of information. More
specifically all of the relevant thermodynamical quantities have
been computed with and without quarks
\cite{Boyd:1996bx,{Okamoto:1999hi}} at zero quark chemical
potential. Very recently new methods have been proposed which
might help studying the phase diagram at non zero chemical
potential via lattice simulations \cite{Fodor:2002hs}. However
still much is left to be understood about the nature of the
transition of hot hadronic matter to a plasma of deconfined quarks
and gluons \cite{Satz:2002mg}.

\section{Dense and Cold QCD: Color Superconductivity}
 At zero temperature but very high quark chemical potential the
perturbative (``naive") vacuum is unstable. In this regime strong
interactions favor the formation of quark-quark condensates in the
color antisymmetric channel \cite{REV}. Possible physical
applications are related to the physics of compact objects
\cite{REV}, supernovae cooling \cite{Carter:2000xf} and explosions
\cite{HHS} as well as to the Gamma Ray Bursts puzzle \cite{OS}.

 The color
superconductive phase is characterized by its gap energy
($\Delta$) associated to quark-quark pairing which leads to the
spontaneous breaking of the color symmetry. According to the
number of light flavors in play we can have different phases.

\subsection{Color Flavor Locked Phase}

     \label{3f}

Let us start with the case of $N_{f}=3$ light flavors. At zero
density only the confined Goldstone phase is allowed and the
resulting symmetry group is $SU_{V}(3)\times U_{V}(1)$.

Turning on low baryon chemical potential we expect the theory to
remain in the confined phase with the same number of Goldstone
bosons (i.e. 8). At very high chemical potential, dynamical
computations suggest that the preferred phase is a superconductive
one and the following ansatz for a quark-quark type of condensate
is energetically favored:
\begin{equation}
\epsilon ^{\alpha \beta }<q_{L\alpha ;a,i}q_{L\beta ;b,j}>\sim
k_{1}\delta _{ai}\delta _{bj}+k_{2}\delta _{aj}\delta _{bi}\ .
\label{condensate}
\end{equation}
\noindent A similar expression holds for the right transforming
fields. The Greek indices represent spin, $a$ and $b$ denote color
while $i$ and $j$ indicate flavor. The condensate breaks the gauge
group completely while locking the left/right transformations with
color. The final global symmetry group is $SU_{c+L+R}(3)$, and the
low energy spectrum consists of $9$ Goldstone bosons.

The low energy effective theory for 3 flavors (CFL) has been
developed in \cite{CG}. We refer to \cite{REV,Casalbuoni,Nardulli}
for a complete summary and review of this phase.

\subsection{2 SC \& Partial Deconfinement}

\label{tre}

QCD with 2 massless flavors has gauge symmetry $SU_{c}(3)$ and
global symmetry
\begin{equation}
SU_{L}(2)\times SU_{R}(2)\times U_{V}(1)\ .
\end{equation}
\noindent At very low baryon chemical potential it is reasonable
to expect that the confined Goldstone phase persists. However at
very high density the ordinary Goldstone phase is no longer
favored compared with a superconductive one associated to the
following type of diquark condensates:
\begin{equation}
\langle L{^{\dagger }}^{a}\rangle \sim \langle \epsilon
^{abc}\epsilon ^{ij}q_{Lb,i}^{\alpha }q_{Lc,j;\alpha }\rangle \
,\qquad \langle R{^{\dagger }}^{a}\rangle \sim -\langle \epsilon
^{abc}\epsilon ^{ij}q_{Rb,i;\dot{\alpha}
}q_{Rc,j}^{\dot{\alpha}}\rangle \ ,
\end{equation}
$q_{Lc,i;\alpha }$, $q_{Rc,i}^{\dot{\alpha}}$ are respectively the
two component left and right spinors. $\alpha ,\dot{\alpha}=1,2$
are spin indices, $c=1,2,3$ stands for color while $i=1,2$
represents the flavor. If parity is not broken spontaneously, we
have $\left\langle L_{a}\right\rangle =\left\langle
R_{a}\right\rangle =f\delta _{a}^{3}$, where we choose the
condensate to be in the 3rd direction of color. The order
parameters are singlets under the $SU_{L}(2)\times SU_{R}(2)$
flavor transformations while possessing baryon charge
$\frac{2}{3}$. The vev leaves invariant the following symmetry
group:
\begin{equation}
\left[ SU_{c}(2)\right] \times SU_{L}(2)\times SU_{R}(2)\times
\widetilde{U}_{V}(1)\ ,
\end{equation}
where $\left[ SU_{c}(2)\right] $ is the unbroken part of the gauge
group. The $\widetilde{U}_{V}(1)$ generator is linear combination
of the previous $U_{V}(1)$ generator and the broken diagonal
generator of the $SU_{c}(3)$ gauge group $T^{8}$. The quarks with
color $1$ and $2$ are neutral under $\widetilde{U}_{V}(1)$ and
consequently the condensate is neutral too.

The superconductive phase for $N_{f}=2$ possesses the same global
symmetry group as the confined Wigner-Weyl phase \cite{S}.  The
effective theory for 2SC can be found in \cite{CDS} while the
effective theories describing the electroweak interactions for the
low-energy excitations in the 2SC and CFL case can be found in
\cite{CDS2001}. The global anomalies matching conditions for 2 and
3 flavors and constraint are discussed in \cite{S}.

An interesting property of the $2SC$ state is that the three color
gauge group breaks via a dynamical higgs mechanism to a left over
$SU_c(2)$ subgroup. In Reference \cite{rischke2k} it has been
shown that the confining scale of the unbroken $SU_c(2)$ color
subgroup is lighter than the superconductive gap $\Delta$. The
confined degrees of freedom, glueball-like particles, are expected
to  be light with respect to $\Delta$, and the effective theory
based on the anomalous variation of the dilation current has been
constructed in \cite{OS2}. Using this model Lagrangian extended to
include non zero temperature in \cite{Sannino:2002re} the
following simple expression for the $SU_c(2)$ critical temperature
was found:
\begin{equation}
T_{c}=\sqrt[4]{\frac{90\,v^3}{2e\pi^2}}\hat{\Lambda} \ .
\label{ApproxTc}
\end{equation}
\begin{figure}[hbtp]
\begin{center}
\epsfig{file=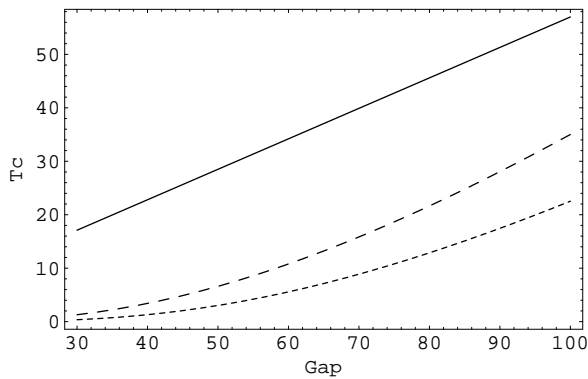, width=8.0cm}
\end {center}
\caption{Plot of the $SU_c(2)$ critical temperature for 2 values
of the quark chemical potential ($\mu=400$~MeV long--dashed line;
$\mu=500$~MeV short--dashed line) as function of the
superconductive gap $\Delta$. The solid line corresponds to the
critical temperature for the superconductive state $0.57 \Delta$.}
\label{su2}
\end{figure}
Here $e$ is the Euler number, $\hat{\Lambda}$ is the confining
scale of the $SU_c(2)$ gluon-dynamics in 2SC and $v$ is the gluon
\cite{rischke2k} as well as light glueball velocity \cite{OS2}.
The deconfining/confining critical temperature is smaller than the
critical temperature $T_{2SC}$ for the superconductive state
itself which is estimated to be $T_{2SC} \approx 0.57~\Delta$ with
$\Delta$ the 2SC gap \cite{PR}. Knowing the explicit dependence of
the $SU_c(2)$ confining temperature on $\mu$ and $\Delta$ directly
affects astrophysical models for compact stars like the one in
Ref.~\cite{OS}. A second order phase transition is also predicted.
Finally in \cite{Sannino:2002re} it is suggested how ordinary
lattice importance sampling techniques can be used to check these
results and constitute, at the same time, the first simulations
testing the high quark chemical potential but small temperature
region of the QCD phase diagram. Interestingly recently it has
also been shown that at high chemical potential the effective
field theory of low energy modes in dense QCD has positive
Euclidean path integral measure \cite{Hong:2002nn}. For
temperatures in the range $T_c < T < T_{2SC}$ the gapped quark
dynamics is no longer negligible. Using the transport theory some
of the quark temperature dependent effects have been investigated
in \cite{Litim:2001je}. It has also been recently argued that a
2SC phase might not appear in compact stars \cite{Alford:2002kj}.

\begin{figure}[h]
\begin{center}
\epsfig{figure=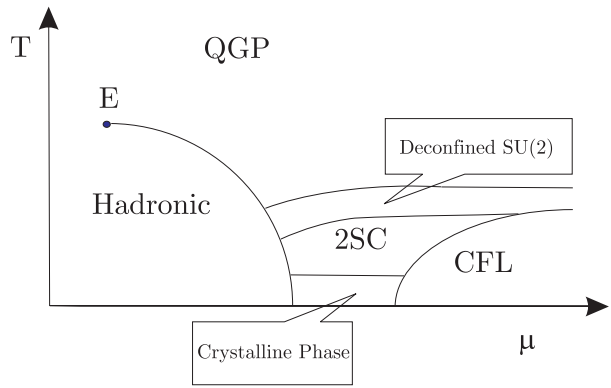,width=8.0cm}\end{center} \caption {An
oversimplified cartoon of the theoretical QCD phase diagram. The
crystalline phase may exists only if we have different chemical
potentials for the up and down quarks. The deconfined $SU_c(2)$
corresponds to the phase where we have gapped up and down quarks
and the remaining unbroken $SU_c(2)$ of color deconfined.}
\label{TChemical}
\end{figure}

In a world with two light and approximately degenerate quarks (the
up and the down) and a relatively heavy strange quark a possible
schematic theoretical representation of the phase diagram is shown
in Fig.~\ref{TChemical}.

\subsection{Crystalline Color Superconductivity/LOFF Phase}
When considering physical applications one cannot neglect the
effects of the quark masses or differences in the chemical
potential among different quark flavors. If the two quarks
participating in the Cooper pair posses different chemical
potentials with the difference denoted by $\delta{\mu}$ a
quark--quark bilinear condensate breaking translational and
rotational invariance may emerge for certain values of $\delta\mu$
\cite{Bowers:2002xr}. A similar phenomenon was suggested in the
context of BCS theory for materials in presence of magnetic
impurities by Larkin, Ovchinnikov, Fulde and Ferrel (LOFF)
\cite{LOFF}. If present in the core of neutron stars an
interesting application of the LOFF phase would be the generation
of glitches \cite{Bowers:2002xr,{Nardulli}}.

\subsection{Kaon Condensation}

The effects of the strange quark mass can be quite dramatic. In
the CFL phase the $K^{+}$ and $K^0$ modes may be unstable for
large values of the strange quark mass signaling the formation of
a kaon condensate \cite{Schafer:2000ew}. Vortex solutions in dense
quark matter due to kaon condensation have been explored in
\cite{Kaplan:uv}.

\section{Higher Spin Condensates}

At non zero quark chemical potential Lorentz invariance is
explicitly broken down to the rotational subgroup $SO(3)$ and
higher spin fields can condense, thus enriching the phase diagram
structure of QCD and QCD-like theories.  The simplest non zero
spin condensate which can appear is the spin one condensate. At
this point we distinguish different types of spin one condensate.

\subsection{Superconductive Vectorial Gaps}
Rotational symmetry can break in a color superconductor if two
quarks of the same flavor gap. In this case the quarks must pair
in a spin one state and a careful analysis has been performed in
\cite{Pisarski:1999tv}. Whether this gap occurs or not in practice
is a very dynamical issue recently investigated in
\cite{Buballa:2002wy}.

\subsection{Vectorial Bose-Einstein type condensation in 2 and 3 color QCD}
Differently from superconductive higher spin gaps, the vectorial
Bose-Einstein type of condensation requires some, already present
in the theory, composite or elementary higher spin states to
couple to an external chemical potential associated to a conserved
current. A simple example is the standard QCD vector field $\rho$
at non zero isospin chemical potential. It is not completely
unexpected that vectors can be relevant at non zero chemical
potential since they already play an important role in QCD at zero
chemical potential \cite{Harada:1995dc} when trying to describe
low energy dynamics.

\begin{figure}[h]
\begin{center}
\epsfig{figure=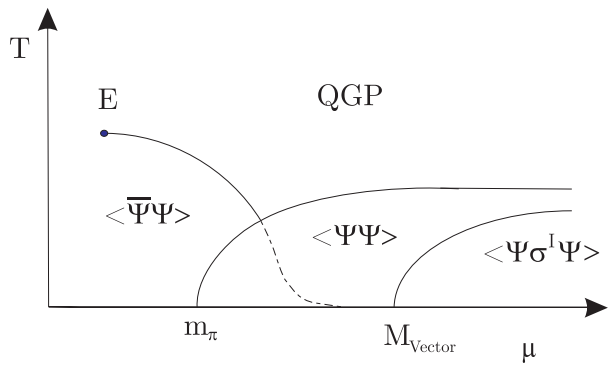, width=8cm}
\end{center}
\caption{Schematic representation of the 2 colors phase diagram.
$m_{\pi}$ is the pion mass and $M_{Vector}$ is the mass of the
lightest vector state. We have chosen a normalization of the
baryon number such that the diquark has unit baryon number.
\label{2colors}}
\end{figure}

Using an effective Lagrangian approach for relativistic vector
fields it was shown  how physical vector fields condense
\cite{Sannino:2001fd} and how the goldstone theorem is modified in
this case \cite{Nielsen}. Vector condensation for strongly
interacting theories has also been suggested in
\cite{Lenaghan:2001sd} in the framework of two color QCD at non
zero quark chemical potential where vector diquark states are
present. A possible phase diagram of two color QCD for given
number of light flavors augmented by a vector condensate phase is
depicted in Fig.~{\ref{2colors}}.

In the third phase on the right we schematically represented the
emergence of the vector condensate. It should coexists with the
diquark condensate \cite{Lenaghan:2001sd}. In this region not only
we have spontaneous breaking of the baryon number but also
breaking of the rotational invariance. The transition is predicted
to be second order \cite{Sannino:2001fd}.

Recent lattice simulations seem to support these predictions
\cite{MPLombardo}. If these results are confirmed then for the
first time we observe spontaneous rotational breaking due solely
to strongly interacting matter. This fact would have far reaching
consequences.

The subject of 2 color QCD at non zero quark chemical potential
deserves a review on its own we just refer here to
\cite{Kogut:2001na,{Nc=2CPT}}.

We can already imagine a number of physical applications of
topical interest. For example in the core of neutron stars or any
compact object featuring a  CFL state vector composite
quasi-particles with masses of the order of the gap can condense
due to the presence of a non zero isospin chemical potential
and/or (as for kaon condensation) due to the effects of a non
small strange mass. This  may lead to the presence of new type of
vortices affecting the compact star dynamics \cite{Kaplan:uv}.

Interestingly vectorial Bose-Einstein condensation recently has
attracted much attention in condensed matter physics since it has
been observed experimentally in alkali atom gases \cite{VBE}.

\section{Phases of Gauge Theories}

To discuss the QCD phase diagram as function of the number of
flavors and colors at zero quark chemical potential and zero
temperature we need to discuss/define in somewhat more details the
phases of gauge theories.

A reasonable way to characterize the phases of gauge theories is
via the potential $V(r)$ between electric test charges separated
by a large distance $r$. The potential (up to a non-universal
additive constant) is conjectured to behave as in the second
column of Table~{\ref{phases}}.
\begin{table}[h]
\begin{center}
\footnotesize
\begin{tabular}{|c|c|c|}
\hline \raisebox{0pt}[13pt][7pt]{\bf Phases} &
\raisebox{0pt}[13pt][7pt]{Electric $V(r)$}
&\raisebox{0pt}[13pt][7pt]{Magnetic $V(r)$}\\
\hline \hline
\raisebox{0pt}[13pt][7pt]{Coulomb}&\raisebox{0pt}[13pt][7pt]{$\frac{1}{r}$}& \raisebox{0pt}[13pt][7pt]{$\frac{1}{r}$}\\
\raisebox{0pt}[13pt][7pt]{Free Electric} &
\raisebox{0pt}[13pt][7pt]{$\frac{1}{r \log{r\Lambda}}$}&
\raisebox{0pt}[13pt][7pt]{$\frac{\log{r\Lambda}}{r}$}
\\ \raisebox{0pt}[13pt][7pt]{Free Magnetic}&\raisebox{0pt}[13pt][7pt]{$\frac{\log{r\Lambda}}{r}$}
& \raisebox{0pt}[13pt][7pt]{$\frac{1}{r\log{r\Lambda}}$}
\\ \raisebox{0pt}[13pt][7pt]{ Higgs}&\raisebox{0pt}[13pt][7pt]{Constant}& \raisebox{0pt}[13pt][7pt]{$\rho r$}
\\  \raisebox{0pt}[13pt][7pt]{Confining}&\raisebox{0pt}[13pt][7pt]{$\sigma {r}$}&  \raisebox{0pt}[13pt][7pt]{Constant}
\\
\hline
\end{tabular}
\end{center}
\caption{Phases of Gauge Theories\label{phases}}
\end{table}
The first three phases have massless gauge fields and potentials
of the form $V(r)\sim e^2(r)/r$.  In the Coulomb phase, the
electric charge $e^2(r)$ is constant.  In the free electric phase,
massless electrically charged fields renormalize the charge to
zero at long distances as $e^{-2}(r)\sim \log(r\Lambda )$. Similar
behavior occurs when the long distance theory is a non-Abelian
theory which is not asymptotically free.  The free magnetic phase
occurs when there are massless magnetic monopoles, which
renormalize the electric coupling constant to infinity at large
distance with a conjectured behavior $e^2(r)\sim \log (r\Lambda
)$. In the Higgs phase, the condensate of an electrically charged
field gives a mass gap to the gauge fields by the Anderson-Higgs
mechanism and screens electric charges, leading to a potential
which, up to the additive non-universal constant, has an
exponential Yukawa decay to zero at long distances.  In the
confining phase, there is a mass gap with electric flux confined
into a thin tube, leading to the linear potential with string
tension $\sigma$.

All of the above phases can be non-Abelian as well as Abelian.  In
particular, in addition to the familiar Abelian Coulomb phase,
there are theories which have a non-Abelian Coulomb phase with
massless interacting quarks and gluons exhibiting the above
Coulomb potential. This phase occurs when there is a non-trivial,
infrared fixed point of the renormalization group.  These are thus
non-trivial, interacting four dimensional conformal field
theories.

It is instructive to consider the behavior of the potential $V(r)$
for magnetic test charges separated by a large distance $r$ (3th
column in Table~{\ref{phases}} ). The behavior in the first three
phases can be written as $V(r)=g^2(r)/r$ where the effective
magnetic charge $g^2(r)$ is related to the effective electric
charge appearing by the Dirac condition, $e(r)g(r)\sim 1$. The
linear potential in the Higgs phase reflects the string tension in
the Meissner effect.

The above behavior is modified when there are matter fields in the
fundamental representation of the gauge group because virtual
pairs can pop up from the vacuum and completely screen the
sources.


Note that under electric-magnetic duality, which exchanges
electrically charged fields with magnetically charged fields, the
behavior in the free electric phase is exchanged with that of the
free magnetic phase.  Mandelstam and 't Hooft suggested that the
Higgs and confining phases are exchanged by duality. Confinement
can then be understood as the dual Meissner effect associated with
a condensate of monopoles \cite{Giacomo:2002qr}. For
supersymmetric theories it is possible to argue that this picture
is correct \cite{IS}. Dualizing a theory in the Coulomb phase, we
remain in the same phase (the behavior of the potential is
unchanged).

Unfortunately the previous classification of the phases of the
gauge theories is not always possible. Besides if we are
interested in the nature of the phase transition itself (by
varying for example the temperature) the previous picture is not
directly applicable. A better description/understanding can be
achieved by introducing gauge invariant order parameters
\cite{Svetitsky:1982gs}.

For theories with global symmetries we can, quite often, construct
simple order parameters. This is the case of the chiral phase
transition where the quark -- antiquark vacuum expectation value
is the order parameter and is also directly related to the
hadronic states of the theory (pions, etc..). Once the order
parameter is identified a mean field theory approach can be
adopted. There are cases, however, where the identification of the
order parameter is not trivial. And even if the order parameter is
found might not be connected straightforwardly to the physical
states of the theory. This is the case of the pure $SU(N_c)$
Yang-Mills theory where the finite temperature gauge invariant
order parameter is the Polyakov loop $\ell$
\cite{Svetitsky:1982gs,{Pisarski:2001pe}}. Here the global
symmetry group is the center $Z_{N_c}$ of  $SU(N_c)$. On the other
side the physical degrees of freedom of any asymptotically free
gauge theory are hadronic states. Using the Yang-Mills trace
anomaly and the exact $Z_N$ symmetry in \cite{Sannino:2002wb}  a
model able to communicate to the hadrons the information carried
by the order parameter was constructed.

\section{The Phase Diagram Along the Flavor Axes}

Having introduced in the previous section the correct terminology
we can now investigate, at zero temperature and quark matter
density, the behavior of QCD as function of number of light
flavors. Besides the academic interest we also hope to use these
theories to model the electroweak symmetry breaking without using
an elementary higgs field \cite{Lane:2002wv}. Indeed the infrared
behavior of gauge theories changes dramatically when changing the
number of light flavors. Predictions resulting from exact
theoretical treatments are, at the moment, only possible for
supersymmetric theories \cite{IS}. Nevertheless for non
supersymmetric QCD theories it is still possible, using different
methods and models, to sketch a plausible phase diagram which can
be tested via lattice simulations. We focus here on zero quark
chemical potential and temperature. We will however suggest how
the dynamics of large number of flavors QCD might influence any
other phase.

\subsection{The Coulomb/Conformal Phase}
 {}For $N_f > \frac{11}{2}N_c$, the
one-loop beta function of QCD changes sign and the theory loses
asymptotic freedom.  The resulting infrared free theory is now in
a non-Abelian QED-like phase in which neither confinement nor
chiral symmetry breaking is expected.  {}For values of $N_f$ near
but below $\frac{11}{2}N_c$, the beta function develops a
perturbative infrared stable fixed point at which the trace of the
energy momentum tensor vanishes exactly and the theory is a
non-Abelian conformal field theory (Coulomb phase). In this phase,
the coupling constant is small on account of the large number of
flavors and so we do not expect any of the global symmetries to
break.  However, as the number of flavors is decreased, the fixed
point becomes nonperturbative and the coupling constant increases
to a critical value above which chiral symmetry breaks
spontaneously. A dynamical scale is generated and conformal
symmetry is lost. The generation of this scale defines the
critical number of massless flavors, i.e. the minimum number of
flavors for which the gauge theory is still conformal and chiral
symmetry is still intact. Below this critical number of flavors,
the theory is expected to confine and the low energy spectrum is
hadronic. This discussion assumes that the conformal and chiral
phase transitions coincide as function of the number of flavors,
but whether or not this is true is still controversial. This is
not the case for supersymmetric QCD (see right panel in
Fig.~\ref{Conformal}) where for a certain range of flavors the
global symmetries do not break but the theory is still assumed to
confine. We will assume here, as corroborated by lattice
simulations for $N_c=3$ \cite{mawhinney} (though more
investigations are needed), that there is in fact a single
conformal/chiral phase transition. Figure~\ref{Conformal} (left
panel) summarizes the possible phase structure for QCD as a
function of the number of light flavors.

\subsection{Conformal Critical Exponents}

By way of a simple model it is possible to describe the
conformal/chiral phase transition \cite{SS}. Since chiral
symmetry, in the present scenario, is linked to the breaking of
the conformal symmetry the model uses the chiral condensate
$\sigma \sim \bar{q} q$ as the order parameter. Trace and axial
anomaly were also used to constrain part of the potential. In the
absence of the quark masses the condensate vanishes exponentially
fast as $N_f$ approaches the critical value. More precisely the
following result for the physical mass $M_{\sigma }$ and $<\sigma
>$ was derived in \cite{SS}:
\begin{equation}
<\sigma >\simeq \left[ \frac{\gamma -1}{2|C|}\right] ^{\frac{1}{2(\gamma -1)}%
}\Lambda \ ,\qquad M_{\sigma }\simeq 2\sqrt{6}|C|\left[ \frac{\gamma -1}{2|C|%
}\right] ^{\frac{1}{2(\gamma -1)}}\Lambda \ .
\end{equation}
Likewise, in the presence of the quark mass term we have
\begin{equation}
\left[ <\sigma >\right] _{\gamma =1}\simeq \left[ \frac{mN_{f}\Lambda }{2|C|}%
\right] ^{\frac{1}{2}}\ ,\qquad \left[ M_{\sigma }\right] _{\gamma
=1}\simeq 2\left[ 2mN_{f}\Lambda \right] ^{\frac{1}{2}}\ .
\end{equation}
Here $\gamma\geq 1$ (in the confined and highly non perturbative
phase)  is the anomalous dimension of the quark mass operator and
$C$ an unknown coefficient.  Thus the order parameter $\sigma $
for $\gamma =1$ vanishes according to the power $1/2$ with the
quark mass in contrast with an ordinary second order phase
transition where the order parameter is expected to vanish
according to the power $1/3$.  We expect that for a vanishing beta
function close to the conformal point the theory becomes conformal
at $\gamma =1$. This fixes the critical value of the number of
flavors to $4~N_c$. For more details and a complete list of
references we refer the reader to the original paper \cite{SS}.
Lattice simulations would certainly provide very useful in
understanding the nature of this conformal phase and associated
phase transition features.

\subsection{Enhanced Symmetries Scenario}
When the number of flavors is just below the critical value, the
theory still exhibits chiral symmetry breaking but is possible
that the vector spectrum changes quite significantly. In Refs.\
\cite{ARS,DRS} it was suggested that a new global symmetry may be
dynamically generated.  This symmetry acts on the massive spectrum
of the theory and is related to the modification of the second
Weinberg spectral function sum rule near the critical number of
flavors \cite{AS}.

If such an enhanced symmetry emerges, the vectors along the broken
generators become degenerate with those along the orthogonal
directions. {}For $N_c=3$ QCD, this corresponds to mass degenerate
vector ($\rho$-type of field) and axial particles even in the
presence of chiral symmetry breaking.

In the enhanced symmetry scenario, the interactions between the
vectors and the Goldstone excitations are suppressed (in the
derivative expansion). The enhanced symmetry scenario imposes very
stringent constraints on the possible form of the negative
intrinsic parity terms (the ones involving the $\epsilon_{\mu \nu
\rho \sigma}$-terms) as well \cite{DRS} \footnote{Nota Bene: In
reference \cite{DRS} the gauging of the Wess-Zumino term needed
for the saturation of the `t Hooft global anomaly matching has
been computed for 2-colors QCD. The effective theories for 2 color
QCD at zero density for the linear and non linear realization
extended, as well, to include the Electroweak theory and to
contain physical vectors states were first constructed in
\cite{ARS,{DRS}} for an arbitrary number of flavors.}

We conjecture the following phase structure before entering the
conformal phase:
\begin{itemize}
\item[$\bullet$]{Approximate local chiral symmetry for small $N_f$}
\item[$\bullet$]{Parity doubling and an extra global symmetry near the critical $N_f$.}
\end{itemize}
At very low number of flavors the vectors can be included in the
low energy effective theory as almost gauge vectors of the chiral
symmetry. By ``almost" we mean that the vector masses are
considered as minimal sources of breaking of the local chiral
symmetry. This hypothesis is at the base of ``vector dominance"
model, it strongly reduces the number of unknown coefficients in
the effective Lagrangian and it has been widely used in the past
for successful phenomenological investigations. Near the critical
$N_f$, according to our conjecture, a new dynamically generated
global symmetry sets in. Due to this new symmetry the vectors are
almost degenerate while chiral symmetry still breaks
spontaneously. In this regime the new global symmetry strongly
constrains the effective Lagrangian theory.

Interestingly enough when extending this theory to model the
spontaneous symmetry breaking sector of the electroweak theory the
major contribution to the $S$-parameter (proportional
schematically to the difference between the square of the mass of
the axial vector and vector $M^2_A - M^2_V$) is protected by the
new enhanced symmetry and leads to phenomenologically viable
technicolor models of the type of the BESS models
\cite{Casalbuoni:2000gn}.

 {}For a fixed, nonzero
chemical potential, the phase structure as the number of light
flavors is increased should be very rich \cite{Schafer:1999fe}.
{}For instance, when $N_f
>\frac{11}{2}N_c$, the theory is no longer asymptotically free and the low
energy theory is simply the QCD Lagrangian. In this regime
perturbation theory is reliable also at low energy scales.

\begin{figure}[htb]
\parbox[t]{5.5cm} {\leavevmode
\epsfig{figure=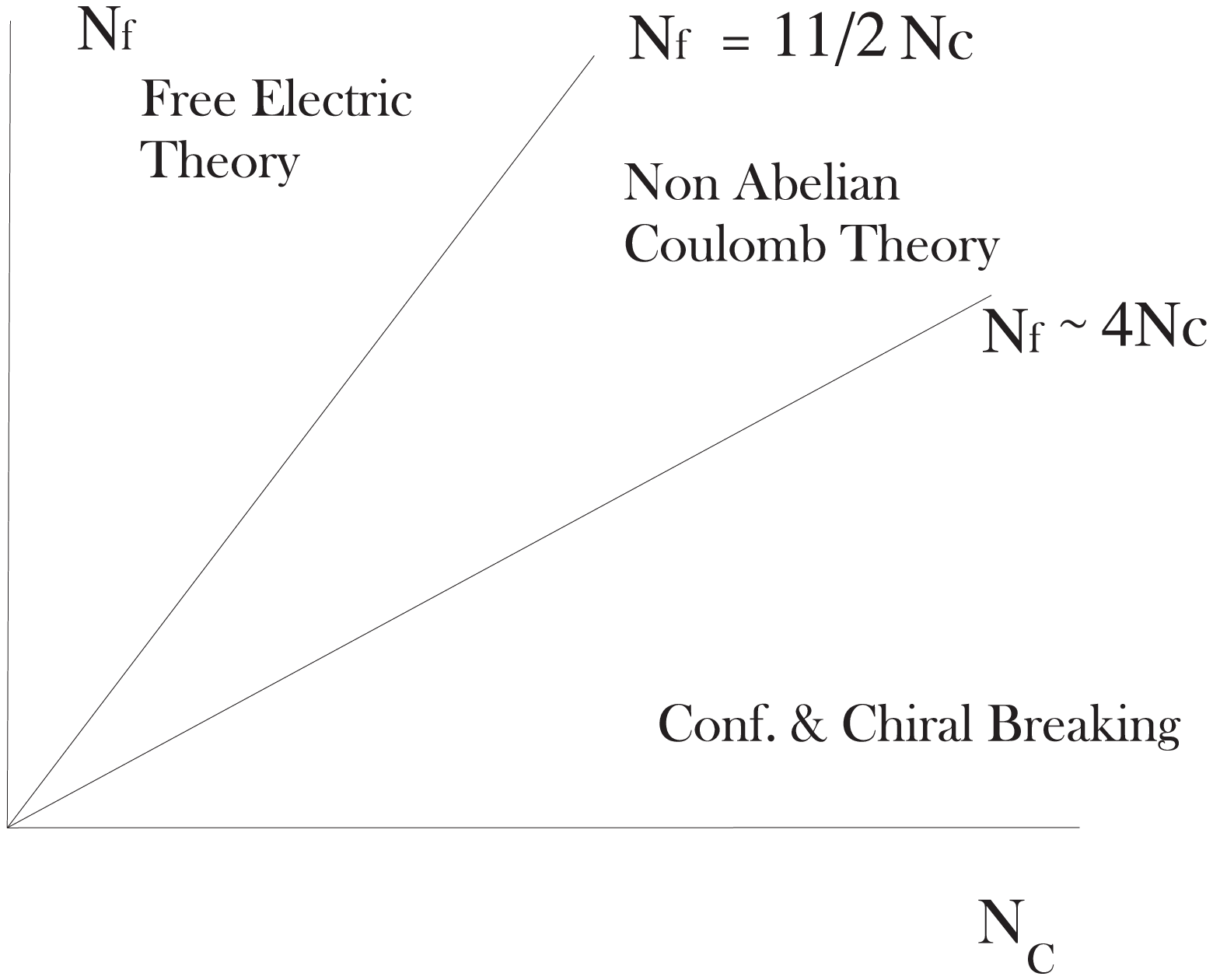,width=5.5cm}} \ \hskip .5cm \
\parbox[t]{5.5cm}
{\leavevmode \epsfig{figure=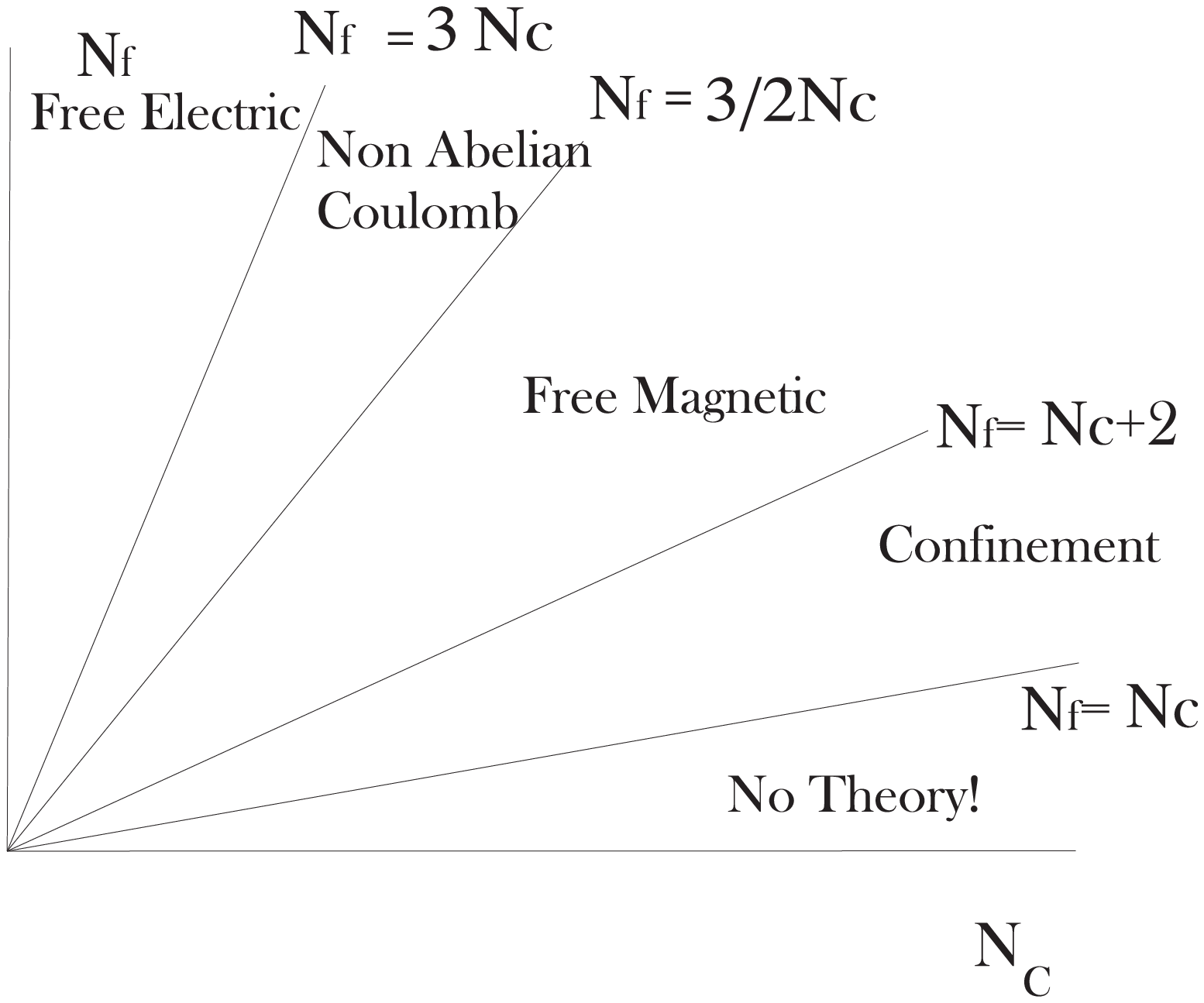,width=5.5cm}}\caption[]
{Left Panel: Represent a possible plan for the QCD phase diagram
at zero quark density and temperature but for different number of
flavors and colors. The origin in both schemes of the number of
color axes corresponds to $N_c=2$.  Right Panel: Same phase
diagram for supersymmetric QCD.} \label{Conformal}
\end{figure}

\section{Conclusions}

The Phase Diagrams in Fig.~{\ref{TChemical}} and in
Fig.~{\ref{Conformal}} are just educated guesses of the true QCD
phase diagram. Indeed other interesting phases may emerge in the
future recalling the richness of the phases encountered in other
physical systems studied in the condensed matter framework.

\section*{Acknowledgments}

It is a pleasure to thank P.~Damgaard, A.D.~Jackson,
K.~Rummukainen, J.~Schechter, W.~Sch\"{a}fer and K.~Splittorff for
discussions and careful reading of the manuscript. I also thank
R.~Casalbuoni, J.T.~Lenaghan, M.P.~Lombardo and G.~Nardulli for
useful discussions. This work is supported by the Marie--Curie
fellowship under contract MCFI-2001-00181.

\end{document}